\documentclass{article}
\usepackage[utf8]{inputenc}

\usepackage{amssymb}
\usepackage{amsmath}
\usepackage{bussproofs}
\usepackage{mathtools}

\usepackage[round]{natbib}
\usepackage{hyperref}

\usepackage{footmisc}

\hypersetup{
     colorlinks   = true, 
     citecolor    = black, 
     linkcolor    = red, 
}

\title{Constructive validity of a generalized Kreisel-Putnam rule}
\author{Ivo Pezlar\footnote{Czech Academy of Sciences, Institute of Philosophy, Jilska 1, Prague 1, 110 00.}}
\date{}

\makeatletter
\DeclareRobustCommand
  \myvdots{\vbox{\baselineskip4\p@ \lineskiplimit\z@
    \hbox{.}\hbox{.}\hbox{.}}}
\makeatother

\begin{document}

\maketitle

\begin{abstract}

In this paper, we propose a computational interpretation of the \textcolor{black}{generalized Kreisel-Putnam rule, also known as the generalized Harrop rule or simply the Split rule}, in the style of BHK semantics. We will achieve this by exploiting the Curry-Howard correspondence between formulas and types. First, we inspect the inferential behavior of the Split rule in the setting of a natural deduction system for the intuitionistic propositional logic. This will guide our process of formulating an appropriate program that would capture the corresponding computational content of the typed Split rule. In other words, we want to find an appropriate selector function for the Split rule by considering its typed variant. Our investigation can also be reframed as an effort to answer the following questions: is the Split rule constructively valid in the sense of BHK semantics? \textcolor{black}{Our answer is positive for the Split rule as well as for its newly proposed generalized version called the S rule.}


\medskip

\end{abstract}

\section{Introduction}

The Harrop rule (\cite{harrop1960}) also known as the Independence of Premise rule or the Kreisel-Putnam rule:

\begin{prooftree}
\AxiomC{$ \neg C \to (A \vee B) $}
\RightLabel{Harrop}
\UnaryInfC{$(\neg C \to A) \vee (\neg C \to B) $}
\end{prooftree}

\noindent is an intriguing rule. It is an admissible but not a derivable rule of intuitionistic logic (\cite{iemhoff2001}), despite being proof-theoretically valid (\cite{piecha2014}) in \textcolor{black}{a variant of} \textcolor{black}{Dummett-}Prawitz-style semantics (\cite{prawitz1971}, \cite{prawitz1973}). If we add it to the intuitionistic logic, we obtain the Kreisel-Putnam logic (\cite{kreisel1957}), which is stronger than the intuitionistic logic yet still has the disjunction property (whenever $A \vee B$ is a theorem, either $A$ or $B$ is a theorem), previously thought to be a property specific to the intuitionistic logic (\cite{lukasiewicz1952}). Furthermore, it is admissible in any intermediate logic (\cite{prucnal1979}).

Yet, its generalized version, which we call the Split rule:\footnote{Where $C$ is a Harrop formula (\cite{harrop1956}), \textcolor{black}{also known as Rasiowa-Harrop formula (\cite{rasiowa1954})}, i.e., a formula in which every disjunction occurs only within the antecedents of implications. Alternatively, borrowing terminology from the literature on \textcolor{black}{iterated} inductive definitions/types, we can say that a Harrop formula is a formula in which every disjunction occurs \emph{negatively} within implications: an occurrence of $X$ in $A \to B$ is \emph{negative} if and only if it is (i) a negative occurrence in $B$ or (ii) a positive occurrence in $A$.}

\begin{prooftree}
\AxiomC{$ C \to (A \vee B) $}
\RightLabel{Split}
\UnaryInfC{$(C \to A) \vee (C \to B) $}
\end{prooftree}

\noindent is arguably even more interesting.\footnote{\textcolor{black}{I} would like to thank V\'{i}t Pun\v{c}och\'{a}\v{r} for bringing this rule and its unexpected appearances in domain logics to \textcolor{black}{my} attention.} If we add it to the intuitionistic logic, we obtain inquisitive \textcolor{black}{intuitionistic} logic (\cite{puncochar2016}, \cite{ciardelli2020}), which has both the disjunctive property and the structural completeness property (enjoyed by classical logic: every admissible rule is derivable), \textcolor{black}{again it can be shown to be proof-theoretically valid in a variant of Dummett-Prawitz-style semantics} (\cite{stafford2021}), yet it is not closed under uniform substitution. Furthermore, it is admissible in any intermediate logic (\cite{minari1988}) and it also makes a surprising appearance in domain logics (\cite{abramsky1991}, \cite{zhang1991}) and we are confident that this list is not complete. 

Despite its significance, the Split rule itself remains mostly unexplored, especially in terms of its proof-theoretic meaning and computational content (a recent exception to this is \cite{condoluci2018} examining the admissibility of the related Harrop rule from the computational view). In this paper, we fill this gap and propose a computational interpretation of the Split rule. We will achieve this by exploiting the Curry-Howard correspondence (\cite{curry1958}, \cite{howard1980}) between formulas and types \textcolor{black}{(also known as the propositions-as-types principle)}. First, we inspect the inferential behavior of the Split rule in the setting of a natural deduction system for the intuitionistic propositional logic. This will then guide our process of formulating an appropriate program that would capture the corresponding computational content of the typed Split rule. In other words, we want to find an appropriate selector function (i.e., a noncanonical eliminatory operator) for the Split rule by considering its typed variant. Our investigation can be thus also reframed as an effort to answer the following questions: is the Split rule constructively valid in the sense of BHK semantics? In other words, 

\begin{itemize}
    \item[] \emph{Can we find a constructive function that would transform arbitrary proofs of the premise of the Split rule into proofs of its conclusion?}
\end{itemize}

Our answer is positive: we propose a selector $\textsf{S}$ for a generalized version of the typed Split rule which we call the S rule. The selector $\textsf{S}$ is based on the selector for the typed disjunction elimination rule and it can be used to justify the non-generalized version as well.

\paragraph{Structure.} This paper is structured as follows: In Section \ref{sec:pre} we briefly introduce the basic concepts presumed by the paper, including the distinction between introduction and elimination rules of natural deduction (\ref{sec:ie}). Readers familiar with the natural deduction framework can skip this section. In Section \ref{sec:generalsplit} we determine the kind of the Split rule with respect to the introduction rules and elimination rules distinction and propose a generalization of the Split rule called the S rule (\ref{sec:split}). \textcolor{black}{Furthemore,} we examine the relationship between the Split rule and the S rule, including their justifications (\ref{sec:splitandS}). And finally in Section \ref{sec:types} we introduce the typed version of the S rule and its corresponding selector $\textsf{S}$.

\section{Preliminaries}
\label{sec:pre}

As a starting point, we choose natural deduction for the intuitionistic propositional logic (IPC) with informal computational semantics based on the BHK interpretation. This choice is motivated by this system's innate affinity towards the formulas-as-types principle, also known as the Curry-Howard correspondence. \textcolor{black}{From Section \ref{sec:types} onwards}, we also adopt \textcolor{black}{Martin-Löf's} constructive type theory, which can be seen as a generalization and formalization of the computational semantics of the intuitionistic logic. \textcolor{black}{We presuppose basic familiarity with these systems on the reader's side. However, to make the presentation as accessible as possible we try to keep the discussion generally informal whenever reasonable.}

The language $\mathcal{L}$ of IPC is a set of formulas $A, B, C, \ldots$ built in the usual way from atomic formulas (propositional variables) $p, q, r, \ldots$ and standard logical constants $\wedge$, $\vee$, $\to$, $\bot$. Negation is defined as $A \to \bot$. 

The meaning of connectives is assumed to be given in terms of canonical proofs, which can be thought of as immediate justifications or direct grounds for asserting formulas\footnote{Strictly speaking, what we are asserting are propositions, not formulas, but for simplicity, we will treat propositions and formulas as interchangeable terms in this paper.} of the corresponding form. The shape of these canonical proofs is prescribed by introduction rules. We discuss this more at length in the next section but, as an example, the meaning of the conjunction $\wedge$ connective is specified by the conjunction introduction rule:

\begin{center}
\AxiomC{$A$}
\AxiomC{$B$}
\RightLabel{$\wedge$I}
\BinaryInfC{$A \wedge B$}
\DisplayProof
\end{center}

\noindent which corresponds to the BHK clause for conjunction: a proof of the formula $A \wedge B$ consists of a proof of $A$ and a proof of $B$. Asserting $A \wedge B$ means that we have proofs of $A$ and $B$ at our disposal. Furthermore, using the formulas-as-types principle, we can make these observations more precise and state that the canonical proof of the formula $A \wedge B$ has the form $(a,b)$ where $a$ is a proof object\footnote{Proof objects (\textcolor{black}{proof terms}, or simply proofs when the meaning is clear from the context), \textcolor{black}{either canonical or noncanonical}, should be thought of as symbolic evidence that something is true and they do not carry epistemic force on their own. The carriers of epistemic force are demonstrations (logical derivations) that proceed from judgment(s) to another judgment.} of $A$ and $b$ is a proof object of $B$. If we are in possession of such proof, we can make the judgment, also called assertion, $(a,b) : A \wedge B$ that informs us that $A \wedge B$ is true. \textcolor{black}{Judgments such as $(a,b) : A \wedge B$ are called categorical judgments as they depend on no assumptions. Judgments depending on assumptions are called hypothetical judgments.}
There are also noncanonical proofs, which can be thought of as delayed justifications or indirect grounds for asserting the corresponding formulas. Noncanonical proofs, however, have to be reducible to the canonical proofs, otherwise, they are meaningless. From the perspective of the formulas-as-types principle, we can view noncanonical proofs as programs and canonical proofs as their values \textcolor{black}{(i.e., as programs that cannot be evaluated/simplified any further)}.

Since by adding Split to IPC we are going beyond the intuitionistic logic, we have to make a few comments regarding the formulas-as-types principle which was initially intended for intuitionistic reasoning only. The standard way of carrying over this principle to stronger logics than the intuitionistic is to simply assume that the new axioms of the stronger logic hold for arbitrary formulas and assign them corresponding proof objects in the spirit of the formulas-as-types principle.

For example, in the case of classical logic, we might assume that the law of excluded middle $A \vee \neg A$ holds for arbitrary formulas and assign it the proof object $\textsf{lem}$, thus obtaining the judgment $\textsf{lem} : A \vee \neg A$. The issue here is, of course, that nobody knows how the proof object $\textsf{lem}$ is supposed to be computed. If we did we would be in a possession of universal decidability procedure for every formula.\footnote{Recall that in the constructive setting the law of excluded middle is not a ``meaningless'' tautology but an judgment of decidability of the proposition $A$.} Thus, the function $\textsf{lem}$ effectively represents an unexecutable black box program.

With the Split rule, we approach the issue analogously. We adopt a new Split axiom schema:

$$(C \to (A \vee B)) \to  ((C \to  A) \vee  (C \to  B))$$

\noindent transform it into a rule form (using its antecedent as the premise and its consequent as the conclusion for the rule) and assign it \textcolor{black}{corresponding proof objects} in the style of the formulas-as-types principle:

\begin{prooftree}
\AxiomC{$f : C \to (A \vee B) $}
\RightLabel{Split}
\UnaryInfC{$\textsf{s} (f) : (C \to A) \vee (C \to B) $}
\end{prooftree}

\noindent \textcolor{black}{where $f$ represents an arbitrary proof object of $C \to (A \vee B)$ and $\textsf{s}$ the selector associated with the Split rule.} The question will then become: how do we compute \textcolor{black}{$\textsf{s}(f)$}? Can we find a general procedure for evaluating the $\textsf{s}$ function (program, selector) and thus expressing the computational content of the corresponding rule? Our answer will be positive but it will require some further generalizations of the Split rule.

\subsection{Introduction and elimination rules}
\label{sec:ie}

The computational content of the natural deduction rules of IPC is closely tied to their inferential behavior which is in turn specified by the introduction and elimination rules (including reduction rules, which link them together) for specific logical connectives.
Generally speaking, an introduction rule for a connective $\circ$ is a rule whose conclusion has $\circ$ as the main connective. An elimination rule is a rule that has this $\circ$-formula as one of its premises.\footnote{See, e.g., \cite{mancosu2021}, p. 69.} 

Semantically speaking, introduction rules ``define'' the meanings of new connectives,\footnote{See \cite{gentzen1935}, \cite{gentzen1969}.} while elimination rules show how to use them. From this perspective, introduction rules are ``self-justifying'' as they effectively act as stipulations.\footnote{See, e.g., \cite{schroeder-heister2006}, p. 532.} On the other hand, elimination rules require justification.\footnote{This would stop to be the case if we switch from the ``verificationist'' approach privileging introduction rules to the ``pragmatist'' approach that views elimination rules as self-justifying and introduction rules as in a need of justification. But as this approach is not the standard one we will not consider it here.} This justification is typically achieved by relating elimination rules to introduction rules via reductions rules: what can be derived from $A$ by elimination rules is to be determined by the premises from which was $A$ canonically derived, i.e., derived by introduction rules for the main operator of $A$.

For example, the implication connective $\to$ is specified by the following implication introduction and elimination rules:

\begin{center}
\AxiomC{$[A]$}
\noLine
\UnaryInfC{$\myvdots$} 
\noLine
\UnaryInfC{$B$}
\RightLabel{\footnotesize $\to$I}
\UnaryInfC{$A \to B$}
\DisplayProof \qquad
\AxiomC{$A \to B$}
\AxiomC{$A$}
\RightLabel{\footnotesize $\to$E}
\BinaryInfC{$B$}
\DisplayProof
\end{center}

\noindent The introduction rule $\to$I tells us what the canonical (meaning constituting) proofs of formulas of the form $A \to B$ should look like. Specifically, it tells us that in order to prove a formula of the form $A \to B$, i.e., \emph{introduce} it into a proof, we first need to find a proof of $B$ from an assumption that we have a proof of $A$, which then can be discharged. In other words, we need to find a procedure that takes an arbitrary proof of $A$ and transforms it into a proof of $B$ (in accordance with the BHK interpretation). The elimination rule $\to$E tells us what can we do with formulas of the form $A \to B$ in proofs. Specifically, it tells us that if we have a proof of $A \to B$ (a major premise) and a proof of $A$ (a minor premise), then we can derive $B$ and, in the process, \emph{eliminate} $A \to B$ from the proof.

Note that the rules $\to$I and $\to$E are, in a way, inverted versions of each other: what goes into introducing $\to$ comes out when eliminating it, no more no less. In other words, if we apply the $\to$I rule and then apply $\to$E immediately after, nothing should be gained or lost in the proof, we are just making an unnecessary detour. 
To check this, consider the following derivation:

\begin{center}
\AxiomC{$[A]$}
\noLine
\UnaryInfC{$\myvdots$}
\noLine
\UnaryInfC{$B$}
\RightLabel{\footnotesize $\to$I}
\UnaryInfC{$A \to B$}
\AxiomC{$A$}
\RightLabel{\footnotesize $\to$E}
\BinaryInfC{$B$}
\DisplayProof
\end{center}

It starts with deriving $B$ (under the active assumption $A$, and possible other assumptions) and ends, again, with deriving $B$. The consecutive applications of $\to$I and $\to$E add no new information: it is just a roundabout way of getting where we started in the first place, i.e., the derivation of $B$.
These detours in derivations can be removed in a process called a detour conversion or a reduction (see \cite{prawitz1965}). We can depict this process via the following ``metarule'' (where $\mathcal{D}$ represents a derivation):

\begin{center}
\AxiomC{$[A]^n$}
\noLine
\UnaryInfC{$\mathcal{D}$}
\noLine
\UnaryInfC{$B$}
\RightLabel{\footnotesize $\to$I$^n$}
\UnaryInfC{$A \to B$}
\AxiomC{$\mathcal{D'}$}
\noLine
\UnaryInfC{$A$}
\RightLabel{\footnotesize $\to$E}
\BinaryInfC{$B$}
\DisplayProof \quad $\xRightarrow[\textrm{$\to$red}]{\textrm{reduces to}}$  \quad
\AxiomC{$\mathcal{D'}$}
\noLine
\UnaryInfC{$A$}
\noLine
\UnaryInfC{$\mathcal{D}$}
\noLine
\UnaryInfC{$B$}
\DisplayProof
\end{center}

\noindent A proof with no detours is called a normal proof or a proof in a normal form.

Analogously, no new information should be gained if we apply introduction rules immediately after elimination rules. Consider, e.g., the following derivation:

\begin{center}
\AxiomC{$A \to B$}
\AxiomC{$[A]$}
\RightLabel{\footnotesize $\to$E}
\BinaryInfC{$B$}
\RightLabel{\footnotesize $\to$I}
\UnaryInfC{$A \to B$}
\DisplayProof
\end{center}

\noindent Similarly as above, it starts with $A \to B$ (and an active assumption $A$) and ends with deriving $A \to B$ (with $A$ discharged). We can depict this process, sometimes called expansion, via the following metarule:

\begin{center}
\AxiomC{$\mathcal{D}$}
\noLine
\UnaryInfC{$A \to B$}
\DisplayProof
\quad $\xRightarrow[\textrm{$\to$exp}]{\textrm{expands to}}$  \quad
\AxiomC{$\mathcal{D}$}
\noLine
\UnaryInfC{$A \to B$}
\AxiomC{$[A]$}
\RightLabel{\footnotesize $\to$E}
\BinaryInfC{$B$}
\RightLabel{\footnotesize $\to$I}
\UnaryInfC{$A \to B$}
\DisplayProof
\end{center}

When introduction and elimination rules behave in this way, i.e., when elimination rules do not allow us to derive more (i.e., they are not too strong = local soundness) or less (i.e., they are not too weak = local completeness) than the introduction rules justify, it is said that they respect the inversion principle (see \cite{prawitz1965}, \cite{lorenzen1955}) or that they are harmonious (see \cite{dummett1991}, \cite{tennant1978}) For a famous example of introduction and elimination rules that are not harmonious, see Prior's Tonk \cite{prior1960}.

Now, let us examine the Split rule.

\section{\textcolor{black}{Generalizing the Split rule}}
\label{sec:generalsplit}
\subsection{The Split rule}
\label{sec:split}

What kind of a rule is the Split rule from the perspective of introduction and elimination rules of natural deduction?\footnote{Recall that the computational content is tied to the inferential content which is then tied to the introduction and elimination rules and their reductions. Thus, we want to position the Split rule within the introduction and elimination rules environment to guide our investigations of its computational content.}

\begin{prooftree}
\AxiomC{$ C \to (A \vee B) $}
\RightLabel{Split}
\UnaryInfC{$(C \to A) \vee (C \to B) $}
\end{prooftree}

\noindent At first look, it seems to be either an introduction rule for the disjunction connective (since $\vee$ appears as the main connective in the conclusion) or an elimination rule for either implication or disjunction, since those are the connectives appearing in the premise.

First, let us consider it as an introduction rule for disjunction. In IPC, the meaning of the $\vee$ connective is already \textcolor{black}{fixed} via its standard introduction rules:

\begin{center}
\AxiomC{$A$}
\RightLabel{$\vee$I$_L$}
\UnaryInfC{$A \vee B$}
\DisplayProof \qquad
\AxiomC{$B$}
\RightLabel{$\vee$I$_R$}
\UnaryInfC{$A \vee B$}
\DisplayProof
\end{center}

\noindent \textcolor{black}{which are taken as fully specifying the meaning of disjunction as a sort of weakening of $A$ or $B$ into $A \vee B$. So, adding a supplementary introduction rule in the form of the Split rule would not only be unwarranted but it would also shift the meaning of disjunction since viewing an inference from $C \to (A \vee B)$ to $(C \to A ) \vee (C \to B)$ as weakening does not seem appropriate.}
Moreover, note that disjunction appears in the premise of the Split rule (i.e., it presupposes we have already introduced $\vee$ and thus understand what it means\footnote{Generally speaking, connective to be defined appearing among the premises does not to be lead to paradoxes (see, e.g., \cite{tranchini2019}, \cite{pezlar2021logica}), since we can have recursive definitions, but this is not the case here.}). Hence, it does not seem reasonable to view the Split rule as a disjunction introduction-like rule.\footnote{When discussing the Split rule, we will talk about introduction-like and elimination-like rules to distinguish them from the standard introduction and elimination rules associated with connectives.}

So, by default, it seems to be an elimination-like rule (or rather \textcolor{black}{more generally, a non-introduction} rule) either for implication or disjunction. 
Both choices are feasible but, for the sake of space, in this paper we will investigate only the second option as it leads to a simpler generalization.\footnote{The first option would require us to additionally consider natural deduction/constructive type theory with higher-level rules.}
Note, however, that if we really want to treat the Split rule as a disjunction elimination-like rule, we need disjunction $\vee$ to be the main connective of its premise, otherwise, it would not fit the general pattern of elimination rules.
\textcolor{black}{To get around this issue, we decompose the original premise of the Split rule into a hypothetical judgment. The resulting rule, built in the style of standard disjunction elimination rule, then looks as follows:}

\begin{prooftree}
\AxiomC{$ [C] $}
\noLine
\UnaryInfC{$\myvdots$}
\noLine
\UnaryInfC{$ A \vee B $}
\AxiomC{$[C \to A]$}
\noLine
\UnaryInfC{$\myvdots$}
\noLine
\UnaryInfC{$D$}
\AxiomC{$[C \to B]$}
\noLine
\UnaryInfC{$\myvdots$}
\noLine
\UnaryInfC{$D$}
\RightLabel{S}
\TrinaryInfC{$D $}
\end{prooftree}

\textcolor{black}{We will call this the S rule and it} can be read as follows: if we derive $A \vee B$ under the assumption $C$ and furthermore we can derive $D$ separately from both $C \to A$ and $C \to B$, then we can proceed to $D$ and discharge the assumptions $C$, $C \to A$, and $C \to B$. 

For now, we will leave the question of justification of this rule open and return to it in Section \ref{sec:types} once we have introduced its typed variant.

\paragraph{Example.} To get a better idea of how the S rule works, let us demonstrate it in practice. Consider the following two formulas:

$$(p \to (q \vee r)) \to ((p \to q) \vee (p \to r))$$

\noindent and 

$$((s \vee t) \to (q \vee r)) \to (((s \vee t) \to q) \vee ((s \vee t) \to r))$$

\noindent With the S rule, we can prove the first formula as follows:

\begin{prooftree}
\small
\AxiomC{$ [p \to (q \vee r)]^2 $}
\AxiomC{$ [p]^1 $}
\RightLabel{$\to$E}
\BinaryInfC{$ q \vee r $}
\AxiomC{$[p \to q]^3$}
\RightLabel{$\vee$I$_L$}
\UnaryInfC{$(p \to q) \vee (p \to r) $}
\AxiomC{$[p \to r]^4$}
\RightLabel{$\vee$I$_R$}
\UnaryInfC{$(p \to q) \vee (p \to r) $}
\RightLabel{S$_{1,3,4}$}
\TrinaryInfC{$(p \to q) \vee (p \to r) $}
\RightLabel{$\to$I$_2$}
\UnaryInfC{$ (p \to (q \vee r)) \to ((p \to q) \vee (p \to r))$}
\end{prooftree}

However, we are not able to prove the second formula, because its proof would require us to assume instead of an atomic $p$, which is a Harrop formula (since every atom is a Harrop formula), a formula $q \vee r$ which is not a Harrop formula. Thus, we cannot apply the S rule, since $C$ is restricted to Harrop formulas only.\footnote{So, systems with the Split rule and/or the S rule are not closed under uniform substitution: note that the second formula is a substitution instance of the first formula with $(s \vee t)$ substituted for $p$.}

\medskip

Thus, there are two approaches we can take towards the Split rule: we can either view it as an implication elimination-like rule (i.e., the Split rule in its original form) or as a disjunction elimination-like rule in the form of the S rule.
Before we examine which variant of the Split rule is more suitable given our ultimate goal of examining its computational content, we first need to make clear an important question: what is the relationship between the Split rule and the S rule?

\subsection{The Split rule and the S rule}
\label{sec:splitandS}

The relationship between the Split rule and the S rule is perhaps best understood as similar to the relationship between elimination rules and general (generalized, parallel) elimination rules. 

What are general elimination rules? Simply put, they are elimination rules following the ``indirect'' pattern of disjunction elimination rule $\vee$E which can be seen as utilizing the principle of proof by induction: to show that an arbitrary $C$ follows from $A \vee B$ it is sufficient to show that it follows from the ``base cases'', i.e., the canonical proofs of $A \vee B$ from $A$ and from $B$. If we apply this style of reasoning to the elimination rules for other connectives (in IPC, namely, conjunction elimination and implication elimination) we obtain corresponding general elimination rules which have a more general form than their standard variants but are logically equivalent.\footnote{See, e.g., \cite{sh2014}. General elimination rules have many useful properties. For example, they ensure the structural correspondence between natural deduction derivations and sequent calculus derivations required for the translation of former to the latter. See \cite{negri2001}.}

For example, the general elimination rules for conjunction $\wedge$ and implication $\to$ are as follows:

\begin{center}
\AxiomC{$A \wedge B$}
\AxiomC{$[A, B]$}
\noLine
\UnaryInfC{$\myvdots$}
\noLine
\UnaryInfC{$C$}
\RightLabel{\footnotesize $\wedge$GE}
\BinaryInfC{$C$}
\DisplayProof \qquad
\AxiomC{$A \to B$}
\AxiomC{$A$}
\AxiomC{$[B]$}
\noLine
\UnaryInfC{$\myvdots$}
\noLine
\UnaryInfC{$C$}
\RightLabel{\footnotesize $\to$GE}
\TrinaryInfC{$C$}
\DisplayProof 
\end{center}

Let us comment briefly on the general elimination rule for implication. We can read it as follows: if we derive $A \to B$, and we derive some further consequences $C$ from $B$ (at that point without knowing whether $A$ is true) and if it turns out that $A$ is indeed true, then we will know that $C$ is true as well and we can derive it (recall that the immediate justification for deriving $A \to B$ is the existence of a derivation of $B$ under the assumption $A$. So if $C$ can be derived from $B$, then it must be already derivable from $A$).

So, returning to the S rule, we can think of it as a generalized version of the Split rule, similarly as $\to$GE is a generalized version of $\to$E:

\begin{center}
\AxiomC{$A \to B$}
\AxiomC{$B$}
\RightLabel{\footnotesize $\to$E}
\BinaryInfC{$B$}
\DisplayProof \quad $\textrm{\small generalizes to}$ \quad
\AxiomC{$A \to B$}
\AxiomC{$A$}
\AxiomC{$[B]$}
\noLine
\UnaryInfC{$\myvdots$}
\noLine
\UnaryInfC{$C$}
\RightLabel{\footnotesize $\to$GE}
\TrinaryInfC{$C$}
\DisplayProof 
\end{center}

\begin{center}
\small 
\AxiomC{$ C \to (A \vee B) $}
\RightLabel{Split}
\UnaryInfC{$(C \to A) \vee (C \to B) $}
\DisplayProof \quad $\textrm{\small generalizes to}$ \quad
\AxiomC{$ [C] $}
\noLine
\UnaryInfC{$\myvdots$}
\noLine
\UnaryInfC{$ A \vee B $}
\AxiomC{$[C \to A]$}
\noLine
\UnaryInfC{$\myvdots$}
\noLine
\UnaryInfC{$D$}
\AxiomC{$[C \to B]$}
\noLine
\UnaryInfC{$\myvdots$}
\noLine
\UnaryInfC{$D$}
\RightLabel{S}
\TrinaryInfC{$D $}
\DisplayProof 
\end{center}

As we have mentioned, elimination rules and their generalized counterparts are equivalent. Does this hold for the Split rule and the S rule as well?
As it turns out, they are indeed equivalent. This can be shown as follows. Suppose that we have at our disposal the S rule and further suppose that we have derived the premises for the Split rule, i.e., we have a derivation (i.e., we know the premise) \AxiomC{$\mathcal{D}_1$}\noLine\UnaryInfC{$C \to (A \vee B)$}\DisplayProof of $C \to (A \vee B)$. Now, assuming $\mathcal{D}_1$ is either a canonical proof of $C \to (A \vee B)$ (i.e., we have immediate justification for asserting it) or that can be reduced to such a proof by a series of finite steps, it is of the form \AxiomC{$[C]$}\noLine\UnaryInfC{$\mathcal{D}_1'$}\noLine\UnaryInfC{$A \vee B$}\RightLabel{$\to$I}\UnaryInfC{$C \to (A \vee B)$}\DisplayProof where \AxiomC{$C$}\noLine\UnaryInfC{$\mathcal{D}_1'$}\noLine\UnaryInfC{$A \vee B$}\DisplayProof is a derivation of $A \vee B$ from $C$. From this derivation we can obtain by means of the S rule (and the $\vee$I rules) a derivation:

\begin{center}
\AxiomC{$ [C] $}
\noLine
\UnaryInfC{$\mathcal{D'}_1$}
\noLine
\UnaryInfC{$ A \vee B $}
\AxiomC{$[C \to A]$}
\RightLabel{$\vee$I$_L$}
\UnaryInfC{$(C \to A) \vee (C \to B)$}
\AxiomC{$[C \to B]$}
\RightLabel{$\vee$I$_R$}
\UnaryInfC{$(C \to A) \vee (C \to B)$}
\RightLabel{S}
\TrinaryInfC{$(C \to A) \vee (C \to B)$}
\DisplayProof 
\end{center}

\noindent which is a derivation of the conclusion $(C \to A) \vee (C \to B)$ of the Split rule.

\medskip

Now, the other direction. Let us suppose that the Split rule is at our disposal and that we have derived the premises of the S rule, i.e., we have a derivation \AxiomC{$C$}\noLine\UnaryInfC{$\mathcal{D}_1$}\noLine\UnaryInfC{$A \vee B$}\DisplayProof of $A \vee B$ from $C$, a derivation \AxiomC{$C \to A$}\noLine\UnaryInfC{$\mathcal{D}_2$}\noLine\UnaryInfC{$D$}\DisplayProof of $D$ from $C \to A$, and a derivation \AxiomC{$C \to B$}\noLine\UnaryInfC{$\mathcal{D}_3$}\noLine\UnaryInfC{$D$}\DisplayProof of $D$ from $C \to B$. Thus, since we have a derivation \AxiomC{$C$}\noLine\UnaryInfC{$\mathcal{D}_1$}\noLine\UnaryInfC{$A \vee B$}\DisplayProof (i.e., the first premise of the S rule), we can further assume that we can obtain a derivation \AxiomC{$\mathcal{D}_1'$}\noLine\UnaryInfC{$C \to (A \vee B)$}\DisplayProof of $C \to (A \vee B)$ via the $\to$I rule.

Now, if we replace every occurrence of the assumption $C \to (A \vee B)$ in derivation $\mathcal{D}_2$ with an application of the Split rule using $\mathcal{D'}_1$ as a premise derivation:

\begin{center}
\AxiomC{$\mathcal{D}_1'$}
\noLine
\UnaryInfC{$C \to (A \vee B)$}
\UnaryInfC{$(C \to A)\vee(C \to B)$}
\DisplayProof
\end{center}

\noindent we obtain a derivation of $D$, i.e., the conclusion of the S rule. Analogously for the other minor premise with derivation $\mathcal{D}_3$.

\subsection{Justification of the Split rule}
\label{sec:SplitJustification}

We have said that introduction rules are typically viewed as self-justifying in contrast to elimination rules that require further justification. Now, since we have decided to view the Split rule as an elimination-like rule it is in need of further justification as well.
As we have mentioned, justification of the elimination rule involves specifying certain reduction procedures for derivations that were derived by elimination rules. For example, the implication elimination rule can be regarded as justified with respect to the reduction procedure $\to$red.

This line of reasoning is another way to understand the connection between the Split rule and the S rule: we can view the latter as a means of justifying the former. More specifically, we can consider the following reduction procedure for justifying the Split rule:\footnote{\textcolor{black}{I thank Antonio Piccolomini d'Aragona for suggesting this variant of the reduction rule.} \textcolor{black}{Note also that} we agree with \cite{schroeder-heister2006} (p. 553) that ``[i]n principle, reductions should be definable for derivation structures ending with any non-introduction inference.''}

\begin{center}
\scriptsize
\AxiomC{$\mathcal{D}$}
\noLine
\UnaryInfC{$ C \to (A \vee B) $}
\RightLabel{Split}
\UnaryInfC{$(C \to A) \vee (C \to B) $}
\DisplayProof \quad $\xRightarrow[\textrm{Split-red}]{\textrm{reduces to}}$ \quad
\AxiomC{$\mathcal{D}$}
\noLine
\UnaryInfC{$ C \to (A \vee B) $}
\AxiomC{$[C]^1$}
\RightLabel{$\to$E}
\BinaryInfC{$ A \vee B $}
\AxiomC{$[C \to A]^2$}
\RightLabel{$\vee$I$_L$}
\UnaryInfC{$(C \to A) \vee (C \to B) $}
\AxiomC{$[C \to B]^3$}
\RightLabel{$\vee$I$_R$}
\UnaryInfC{$(C \to A) \vee (C \to B) $}
\RightLabel{S$_{1,2,3}$}
\TrinaryInfC{$(C \to A) \vee (C \to B) $}
\DisplayProof 
\end{center}

\noindent Furthermore, note that this reduction is distinct from the standard reductions as it relies on ``external'' rules. Namely, disjunction introduction rules, which are trivially justified as introduction rules, \textcolor{black}{implication elimination rule (which is justified by its corresponding reduction rule $\to$red)}, and the S rule. Compare this with, e.g., the standard reduction procedure for implication $\to$red which relies only on given subderivations and it invokes no other rules. 

We can see that this justification of the Split rule relies, among other rules, on the S rule. The S rule -- as an elimination-like rule itself -- is in a need of justification as well. Thus, we will also need to supply reduction procedures for it.
Since we treat the S rule as a disjunction elimination-like rule, the notion of reduction/detour conversion still makes sense, as it is possible to introduce the disjunction $A \vee B$ under the assumption $C$ and then immediately eliminate it via the S rule, which then constitutes an unnecessary detour in a derivation (analogously with the Split rule). The reduction rules we obtain are as follows:

\begin{center}
\small 
\AxiomC{$\textcolor{black}{[ C ]}$}
\noLine
\UnaryInfC{$\mathcal{D}_1$}
\noLine
\UnaryInfC{$ A $}
\RightLabel{$\vee$I$_L$}
\UnaryInfC{$ A \vee B $}
\AxiomC{$[C \to A]$}
\noLine
\UnaryInfC{$\mathcal{D}_2$}
\noLine
\UnaryInfC{$D$}
\AxiomC{$[C \to B]$}
\noLine
\UnaryInfC{$\mathcal{D}_3$}
\noLine
\UnaryInfC{$D$}
\RightLabel{S}
\TrinaryInfC{$D $}
\DisplayProof \quad $\xRightarrow[\textrm{S-red$_L$}]{\textrm{reduces to}}$  \quad
\AxiomC{$\textcolor{black}{ [C] }$}
\noLine
\UnaryInfC{$\mathcal{D}_1 $}
\noLine
\UnaryInfC{$ A $}
\noLine
\UnaryInfC{$\mathcal{D}_{2}$}
\noLine
\UnaryInfC{$D$}
\DisplayProof 
\end{center}

\smallskip

\begin{center}
\small
\AxiomC{$\textcolor{black}{ [ C ] }$}
\noLine
\UnaryInfC{$\mathcal{D}_1$}
\noLine
\UnaryInfC{$ B $}
\RightLabel{$\vee$I$_R$}
\UnaryInfC{$ A \vee B $}
\AxiomC{$[C \to A]$}
\noLine
\UnaryInfC{$\mathcal{D}_2$}
\noLine
\UnaryInfC{$D$}
\AxiomC{$[C \to B]$}
\noLine
\UnaryInfC{$\mathcal{D}_3$}
\noLine
\UnaryInfC{$D$}
\RightLabel{S}
\TrinaryInfC{$D $}
\DisplayProof \quad $\xRightarrow[\textrm{S-red$_R$}]{\textrm{reduces to}}$ \quad
\AxiomC{$\textcolor{black}{ [C] }$}
\noLine
\UnaryInfC{$\mathcal{D}_1 $}
\noLine
\UnaryInfC{$B$}
\noLine
\UnaryInfC{$\mathcal{D}_{3}$}
\noLine
\UnaryInfC{$D$}
\DisplayProof 
\end{center}

\noindent With the reductions S-red$_L$ and S-red$_R$ the justification of the Split rule is achieved (we will inspect these reductions, more specifically, their typed variants in the form of computations rules more in the next section).

\section{Formulas as types: Typing the S rule}
\label{sec:types}

Let us return to our original task, i.e., investigating the computational content of the Split rule in the style of BHK semantics and guided by the formulas-as-types principle while assuming a propositional fragment of Martin-L\"{o}f's constructive type theory on the background. 

As mentioned, we will regard the Split rule as a disjunction elimination-like rule that can be generalized into the S rule.
Thus, let us start by considering the standard typed variant of the disjunction elimination rule before we try to model the selector for S based upon it.

The typed introduction and elimination rules specifying the constructors and the selector for disjunction are the following:

\begin{center}
\small 
\AxiomC{$a : A$}
\RightLabel{$\vee$I$_L$}
\UnaryInfC{$\textsf{inl}(a) : A \vee B$}
\DisplayProof
\quad
\AxiomC{$b : B$}
\RightLabel{$\vee$I$_L$}
\UnaryInfC{$\textsf{inr}(b) : A \vee B$}
\DisplayProof
\end{center}

\begin{center}
\small 
\AxiomC{$c : A \vee B $}
\AxiomC{$[x : A]$}
\noLine
\UnaryInfC{$\myvdots$}
\noLine
\UnaryInfC{$d(x) : D$}
\AxiomC{$[y : B]$}
\noLine
\UnaryInfC{$\myvdots$}
\noLine
\UnaryInfC{$e(y) : D$}
\RightLabel{$\vee$E}
\TrinaryInfC{$ \textsf{D}(c, d, e) : D $}
\DisplayProof
\end{center}

The constructors $\textsf{inl}$ and $\textsf{inr}$ for the type $A \vee B$ are called injections and they tell us from which disjunct was the disjunction constructed. The selector $\textsf{D}$ is a function that takes three arguments (an arbitrary proof of $A \vee B$, a function $d$ that transforms an arbitrary proof of $A$ into a proof of $D$, and a function $e$ that transforms an arbitrary proof of $B$ into a proof of $D$) and returns a proof of $D$ as a value.

Note that the selector $\textsf{D}$  operates essentially as a pattern matching program that incorporates the method of proof by cases by its ability to generate subproofs: it checks whether $A \vee B$ was derived from $A$ or from $B$ (i.e., whether the forms of its canonical proof are $\textsf{inl}(a)$ or $\textsf{inr}(b)$): if from $A$, then we should continue by computing the subprogram  $d$, if it was derived from $B$, then we should continue by computing the subprogram $e$. The corresponding computation rules for these two cases are as follows: 

\begin{center}
\small
\AxiomC{$a : A$}
\AxiomC{$[x : A]$}
\noLine
\UnaryInfC{$\myvdots$}
\noLine
\UnaryInfC{$d(x) : D$}
\AxiomC{$[y : B]$}
\noLine
\UnaryInfC{$\myvdots$}
\noLine
\UnaryInfC{$e(y) : D$}
\RightLabel{$\vee$C$_L$}
\TrinaryInfC{$ \textsf{D}(\textsf{inl}(a), d, e) = d(a) : D $}
\DisplayProof
\quad
\AxiomC{$b : B$}
\AxiomC{$[x : A]$}
\noLine
\UnaryInfC{$\myvdots$}
\noLine
\UnaryInfC{$d(x) : D$}
\AxiomC{$[y : B]$}
\noLine
\UnaryInfC{$\myvdots$}
\noLine
\UnaryInfC{$e(y) : D$}
\RightLabel{$\vee$C$_L$}
\TrinaryInfC{$ \textsf{D}(\textsf{inr}(b), d, e) = e(b) : D $}
\DisplayProof
\end{center}

Now, let us return to the S rule.
If we want to produce a typed variant of the S rule in the style of the typed disjunction elimination rule, we would need to find a program (a three-argument function), let us call it $\textsf{S}$, that would incorporate the method of proof by cases and could bind variables (i.e., it could discharge assumptions).

Immediately, we can see that the selector $\textsf{D}$ seems as a good basis for the typed S rule. It appears to fit all the requirements: it takes three arguments, it can discharge assumptions, but most importantly, the core mechanism of the selector $\textsf{D}$ is subproof generation (specifically, case analysis), not just substitution. And, as it turns out, that is exactly what we need to express the computational content of the S rule, and thus effectively also of the Split rule.

Let us produce the typed variant of the S rule taking all these considerations into account, which will give us the new selector $\textsf{S}$:\footnote{We will refrain from calling the selector $\textsf{S}$ ``split'' as in the literature on type theory a selector named $\textsf{split}$ already appears but it is used as the selector for conjunction, or more precisely, for the cartesian product of two types (see, e.g., \cite{nordstrom1990}, p. 73).}

\begin{prooftree}
\AxiomC{$ [z : C] $}
\noLine
\UnaryInfC{$\myvdots$}
\noLine
\UnaryInfC{$c(z) : A \vee B $}
\AxiomC{$[x : C \to A]$}
\noLine
\UnaryInfC{$\myvdots$}
\noLine
\UnaryInfC{$d(x) : D$}
\AxiomC{$[y : C \to B]$}
\noLine
\UnaryInfC{$\myvdots$}
\noLine
\UnaryInfC{$e(y) : D$}
\RightLabel{S}
\TrinaryInfC{$ \textsf{S}(c, d, e) : D $}
\end{prooftree}

Note that this rule differs from the typed disjunction elimination rule in three key aspects: the first premise is a hypothetical judgment {\small \AxiomC{$ z : C $}\noLine \UnaryInfC{$\myvdots$} \noLine\UnaryInfC{$c(z) : A \vee B $}\DisplayProof}, \textcolor{black}{i.e., $z : C \vdash c(z) : A \vee B$ in linear notation}, the assumptions of the subproofs take the form of an implication (i.e., they are not composed of subformulas of the original disjunction), and the formula $C$ has to be a Harrop formula.

\textcolor{black}{Now, before we can get to the explanation of how to compute the selector $\textsf{S}$, we first need to make a short detour and explain the form of the major premise of the S rule, i.e., the hypothetical judgment $z : C \vdash c(z) : A \vee B$. In Martin-L\"{o}f's constructive type theory, the standard meaning of the hypothetical judgment of the general form: $$x : A \vdash b(x) : B(x) $$ is that $b(a)$ is a proof object for $B(a)$ assuming we have a proof object $a$ for $A$. In other words, its meaning is explained via reducing it to the corresponding categorical judgment:
$$b(a) : B(a)$$
And the meaning explanation of this judgment is that $b(a)$ is a program that upon computation yields a canonical proof object of the type $B(a)$. Note that $b(a) : B(a)$ is obtained by substituting the closed proof object $a$ of the type $A$ for the free variable $x$ in $b(x)$ and $B(x)$.}\footnote{\textcolor{black}{Furthermore, we also need to know that $b(x)$ is extensional in the sense that if $a = a' : A$, then $b(a) = b(a') : A(a')$. See \cite{martin-lof1984}.}}
\textcolor{black}{To put it differently, in order to be able to compute the open proof object $b(x)$ depending on $x : A$ to obtain the canonical proof object of the type $B(x)$, we first need to replace the free variable $x$ with the appropriate closed proof object $a$. Now, let us return to the S rule.}

\textcolor{black}{Observe that the major premise of the S rule is a more specific hypothetical judgment than the one presented above as its assumption is restricted to Harrop formulas only.}
\textcolor{black}{This fact, together with \cite{smith1993}'s results showing us that we can consider open proof objects computable to a canonical form as long as they range over Harrop formulas,\footnote{For details, see Appendix \ref{sec:slash}.} allows us to introduce a specialized variant of the hypothetical judgment of the form:
$$z : C \vdash b(z) : B(z) $$
where $C$ is restricted to Harrop formulas with the following modified meaning explanation: $b(z)$ is a program that upon computation yields a canonical proof object of the type $B(z)$.\footnote{I thank Ansten Klev for this suggestion.} In other words, this hypothetical judgment behaves essentially as a categorical one since $z : C$ is a computationally irrelevant assumption and as such it is not needed for the evaluation of $b(z) : B(z)$.\footnote{It seems to correspond to proof irrelevant assumptions of the form $x \div A$ introduced in \cite{pfenning2001b}, however, for the sake of space, we will not explore this connection further here.} Also, note that this meaning explanation mirrors the meaning explanation of the corresponding categorical judgment, so it does not break the general idea of explaining hypothetical judgments via categorical ones.}

\textcolor{black}{With this specialized hypothetical judgment, we can now explain the computational interpretation of the $\mathsf{S}$ selector.}
So, how do we compute this new program (i.e., noncanonical proof object) of the form $\textsf{S}(c, d, e) $? We begin by computing $c$ to the canonical form. 
\textcolor{black}{First, note that $c$ is actually an open term $c(z)$, i.e., it depends on the variable $z$. Computing a program with a hole might seem odd at first, however, we have to keep in mind that it is not just any hole: the variable $z$ is of type $C$ which is restricted to Harrop formulas only. This makes it an instance of the special kind of hypothetical judgment we have just introduced above and, consequently, it means that $c(z)$ can be computed to a canonical form as is.}
If the value of $c(z)$ is of the form $\textsf{inl}(a(z))$ with $a : A$ and $z : C$ then, lambda abstract over the variable $z$ to obtain $\lambda z . a(z)$ (of type $C \to A$) and continue by computing $d(\lambda z . a(z))$ of type $D$.
If the value of $c(z)$ is of the form $\textsf{inr}(b(z))$ with $b : B$ and $z : C$ then, lambda abstract over the variable $z$ to obtain $\lambda z . b(z)$ (of type $C \to B$) and continue by computing $e(\lambda z . b(z)$ of type $D$.

This explanation of $\mathsf{S}$ is expressed by the following computation rules:

\begin{center}
\scriptsize
\AxiomC{$ [z : C] $}
\noLine
\UnaryInfC{$\myvdots$}
\noLine
\UnaryInfC{$a(z) : A  $}
\AxiomC{$[x : C \to A]$}
\noLine
\UnaryInfC{$\myvdots$}
\noLine
\UnaryInfC{$d(x) : D$}
\AxiomC{$[y : C \to B]$}
\noLine
\UnaryInfC{$\myvdots$}
\noLine
\UnaryInfC{$e(y) : D$}
\TrinaryInfC{$ \textsf{S}(\textsf{inl}(a(z)), d, e) = d(\lambda z.a(z)) : D $}
\DisplayProof
\AxiomC{$ [z : C] $}
\noLine
\UnaryInfC{$\myvdots$}
\noLine
\UnaryInfC{$b(z) : B  $}
\AxiomC{$[x : C \to A]$}
\noLine
\UnaryInfC{$\myvdots$}
\noLine
\UnaryInfC{$d(x) : D$}
\AxiomC{$[y : C \to B]$}
\noLine
\UnaryInfC{$\myvdots$}
\noLine
\UnaryInfC{$e(y) : D$}
\TrinaryInfC{$ \textsf{S}(\textsf{inr}(b(z)), d, e) = e(\lambda z.b(z)) : D $}
\DisplayProof
\end{center}

We can observe that the selector $\textsf{S}$ behaves analogously to the selector $\textsf{D}$: the main difference is that the function $d(x)$ requires an argument of type $C \to A$, not $A$ as with $\textsf{D}$, so instead of substituting $a$ for $x$ in $d(x)$ we substitute $\lambda z . a (z)$ for $x$ in $d(x)$. Also, note that the selector $\textsf{S}$ binds the variable $z$ in $a(z)$. Analogously for the function $e(y)$.

\paragraph{Note.} 
We can regard the typed S rule as a generalized version of the typed disjunction elimination rule. In other words,  we can view the selector $\textsf{D}$ as a special case of the selector $\textsf{S}$: by choosing $c(z)$ to be $\textsf{inl}(a)$ or $\textsf{inr}(b)$ we are making the derivation of $A \vee B$ independent of the assumption $z : C$, which is then correspondingly reflected in the assumptions of the subproofs (recall that they were justified on the ground that $A \vee B$ was derived from $C$, if it wasn't, we have no reason to make these assumptions). Thus, we get $A$ and $B$ instead of $C \to A$ and $C \to B$ since $A \vee B$ no longer depends on $C$. The computation rules will be changed accordingly.

\medskip

Since we have found a function that transforms the premises of the rule S into its conclusion, we can say that the rule is constructively valid.\footnote{We should not conflate the notions of constructive validity, schematic validity (\cite{aragona2023}), and proof-theoretic validity (see \cite{schroeder-heister2006}), which in turn should not be conflated with admissibility. These notions are no doubt related but distinct (see also \cite{sanz2014}).} 
\textcolor{black}{But what about the original Split rule itself? Is it also constructively valid? First, note that we couldn't find a selector for the Split rule, i.e., a constructive function that would take arbitrary proofs of the premise $C \to (A \vee B)$ and transform them into proofs of the conclusion $(C \to A)\vee(C \to B)$. Also, we cannot simply replace the black box placeholder selector $\textsf{s}$ discussed at the beginning:}

\begin{prooftree}
\AxiomC{$f : C \to (A \vee B) $}
\UnaryInfC{$\textsf{s} (f) : (C \to A) \vee (C \to B) $}
\end{prooftree}

\noindent \textcolor{black}{with the selector $\textsf{S}$:}

\begin{prooftree}
\AxiomC{$f : C \to (A \vee B) $}
\UnaryInfC{$\textsf{S} (f) : (C \to A) \vee (C \to B) $}
\end{prooftree}

\noindent \textcolor{black}{and expect it to work. The reason for that is that $\textsf{S}$ takes different arguments than the above derivation provides, namely $\textsf{S}$ is a function that takes three arguments (a function $z.c$ that transforms an arbitrary proof of $C$ into a proof of $A \vee B$, a function $x.d$ that transforms an arbitrary proof of $C \to A$ into a proof of $D$, and a function $y.e$ that transforms an arbitrary proof of $C \to B$ into a proof of $D$) and returns a proof of $D$ as a value. On the other hand, $\textsf{s}$ would have to be a function that requires a single argument (an arbitrary proof $f$ of $C \to (A \vee B)$) and returns a proof of $ (C \to A) \vee (C \to B)$ as a value.}

\textcolor{black}{However, since the Split rule and the S rule are interderivable (see Section \ref{sec:splitandS}), we can justify its constructive validity in an indirect way.\footnote{I thank Antonio Piccolomini d'Aragona for this observation for raising the issue discussed in the following note.}}
\textcolor{black}{We simply apply the Split-red reduction rule (now typed) to a derivation ending with the Split rule (assuming $\mathcal{D}$ is closed valid derivation):}

\begin{center}
\def\ScoreOverhang{1pt} 
\tiny
\AxiomC{$\mathcal{D}$}
\noLine
\UnaryInfC{$f : C \to (A \vee B) $}
\RightLabel{Split}
\UnaryInfC{$\textsf{s}(f) : (C \to A) \vee (C \to B) $}
\DisplayProof \quad $\xRightarrow[\textrm{Split-red}]{\textrm{reduces to}}$ \quad \medskip
\AxiomC{$\mathcal{D}$}
\noLine
\UnaryInfC{$ f : C \to (A \vee B) $}
\AxiomC{$[z : C]^1$}
\RightLabel{$\to$E}
\BinaryInfC{$ \textsf{ap}(f, z) : A \vee B $}
\AxiomC{$[x : C \to A]^2$}
\RightLabel{$\vee$I$_L$}
\UnaryInfC{$\textsf{inl}(x) : (C \to A) \vee (C \to B) $}
\AxiomC{$[y : C \to B]^3$}
\RightLabel{$\vee$I$_R$}
\UnaryInfC{$\textsf{inr}(y) : (C \to A) \vee (C \to B) $}
\RightLabel{S$_{1,2,3}$}
\TrinaryInfC{$ \textsf{S}(\textsf{ap}(f, z) , \textsf{inl}(x), \textsf{inr}(y) ) : (C \to A) \vee (C \to B) $}
\DisplayProof 
\end{center}

\noindent \textcolor{black}{Now, since we know that the S rule is valid, we can claim that this derivation is also valid (assuming $\mathcal{D}$ is valid), and use this piece of knowledge to further justify the claim that the Split rule is valid as well.}
\textcolor{black}{Furthermore, it can be shown that if we add the \textsf{S} selector, i.e., the typed S rule, to Martin-L\"{o}f's constructive type theory, it retains normalization (see Appendix \ref{sec:normal}).}

\medskip

\noindent \textcolor{black}{\textbf{Note}. Could perhaps a simpler justification be found for the Split rule? First, let us consider the following (untyped) derivation containing an application of the Split rule:}

\begin{center}
\AxiomC{$[C]^1$}
\noLine
\UnaryInfC{$\textcolor{black}{\mathcal{D}'}$}
\noLine
\UnaryInfC{$A$}
\RightLabel{$\vee$I$_L$}
\UnaryInfC{$ A \vee B$}
\RightLabel{$\to$I$_1$}
\UnaryInfC{$C \to A \vee B$}
\RightLabel{Split}
\UnaryInfC{$(C \to A) \vee (C \to B) $}
\DisplayProof 
\end{center}

\noindent \textcolor{black}{An analogous derivation can be provided for the right disjunct $B$ but we will now focus only the left disjunct $A$. Now, consider the following reduction procedure (compare with Split-red):}

\begin{center}
\small
\AxiomC{$[C]^1$}
\noLine
\UnaryInfC{$\textcolor{black}{\mathcal{D}'}$}
\noLine
\UnaryInfC{$A$}
\RightLabel{$\vee$I$_L$}
\UnaryInfC{$ A \vee B$}
\RightLabel{$\to$I$_1$}
\UnaryInfC{$C \to A \vee B$}
\RightLabel{Split}
\UnaryInfC{$(C \to A) \vee (C \to B) $}
\DisplayProof $\xRightarrow[\textrm{Split-red2}]{\textrm{reduces to}}$ 
\AxiomC{$[C]^1$}
\noLine
\UnaryInfC{$\textcolor{black}{\mathcal{D}'}$}
\noLine
\UnaryInfC{$A$}
\RightLabel{$\to$I$_1$}
\UnaryInfC{$C \to A $}
\RightLabel{$\vee$I$_L$}
\UnaryInfC{$(C \to A) \vee (C \to B) $}
\DisplayProof 
\end{center}

\noindent \textcolor{black}{Can this be regarded as a simpler, more basic justification of the Split rule since it makes no use of the S rule and only relies on $\vee$I and $\to$I rules? We believe so, however, with one important caveat -- it rather shows that Split is proof-theoretically valid, not necessarily constructively valid. In other words, the above reduction procedure still gives us no indication as to how the corresponding selector function that transforms proofs of the premise of the Split rule into proofs of the conclusion should look like. This becomes more clear when we consider the typed variants:}

\begin{center}
\scriptsize
\AxiomC{$[z : C]^1$}
\noLine
\UnaryInfC{$\textcolor{black}{\mathcal{D}'}$}
\noLine
\UnaryInfC{$a(z) : A$}
\RightLabel{$\vee$I$_L$}
\UnaryInfC{$ \textsf{inl}(a(z)) : A \vee B$}
\RightLabel{$\to$I$_1$}
\UnaryInfC{$ \lambda z . \textsf{inl}(a(z)) : C \to A \vee B$}
\RightLabel{Split}
\UnaryInfC{$\textsf{s}(\lambda z . \textsf{inl}(a(z))) : (C \to A) \vee (C \to B) $}
\DisplayProof $\xRightarrow[\textrm{Split-red2}]{\textrm{reduces to}}$ 
\AxiomC{$[z : C]^1$}
\noLine
\UnaryInfC{$\textcolor{black}{\mathcal{D}'}$}
\noLine
\UnaryInfC{$a(z) : A$}
\RightLabel{$\to$I$_1$}
\UnaryInfC{$ \lambda z . a(z) : C \to A $}
\RightLabel{$\vee$I$_L$}
\UnaryInfC{$ \textsf{inl}(\lambda z . a(z)) : (C \to A) \vee (C \to B) $}
\DisplayProof 
\end{center}

\noindent \textcolor{black}{The question that still remains open is how the function $\textsf{s}$ should be computed. As mentioned above, we cannot simply replace it with $\textsf{s}$. And supplying some straightforward computational rule for $\textsf{s}$ such as $\textsf{s}(\lambda z . \textsf{inl}(a(z))) = \textsf{inl}(\lambda z . a(z)) : (C \to A) \vee (C \to B)$ would also not suffice as we would still need to take care of the other disjunct as well. And if we chose to incorporate it all into a single selector (which would have to combine the mechanisms of assumption withdrawing, substitution, and case analysis), we would end up with the selector $\textsf{S}$ again or some variant of it.}

\section{Conclusion}

We have presented a generalized version of the Kreisel-Putnam rule, also known as the Split rule, called the S rule, and shown that is constructively valid in the sense of BHK semantics. Specifically, we have found an effective function that transforms arbitrary proofs of the premises into proofs of the conclusion. We have called this function the selector $\mathsf{S}$ and it can be used to indirectly justify the Split rule itself. 
The most tricky part of the typed S rule/selector $\mathsf{S}$ lies in the fact that it requires an evaluation of an open proof object to a canonical form. This issue can be, however, overcome once we realize that the free variables of the open proof object range only over Harrop formulas which are computationally irrelevant. Furthermore, we have checked that if we add this rule into constructive type theory, it retains normalization.

Concerning future work, the next natural step would be to explore the other possible generalization of the Split rule, i.e., treat it as an implication elimination-like rule instead of disjunction elimination-like rule and compare the resulting rule with the typed S rule and its selector $\mathsf{S}$.

Concerning related work, \cite{condoluci2018} proposed a typed rule for the (non-generalized) Harrop rule that follows the same general pattern as our typed Split rule (i.e., it is based on the typed disjunction elimination rule). Interestingly, they arrived at this pattern differently: while our approach is bottom-up (we have started by studying the inferential behavior of the Split rule and then generalized it), their approach was top-down (they have started with Visser rules (\cite{roziere1993}, \cite{iemhoff2005}) as a basis for all admissible rules and considered the Harrop rule as a special case). Furthermore, although they also rely on the propositions-as-types principle in their investigation, their goal was different from ours. They were interested in examining the notion of admissibility, while we are interested in the proof-theoretic/computational meaning and constructive validity of the Split rule itself.

\appendix

\newpage

\section{\textcolor{black}{Appendix: Open proof objects computable to canonical forms}}
\label{sec:slash}

\textcolor{black}{Most of the work necessary to show that we can compute open proof objects ranging over Harrop formulas to canonical values has been already done and can be divided into three main observations.}

\begin{itemize}
    \item[] \textcolor{black}{\textbf{Observation 1}. Harrop formulas have no computational/constructive content due to the fact that disjunction can never appear as the main connective.}\footnote{In this paper, we are solely interested in the variant of the Split rule where $C$ is a Harrop formula, however, it might be interesting to consider also different variants. For example, a variant where $C$ is an almost negative formula (\cite{troelstra1973}), or a normal formula (\cite{nepeivoda1978}, \cite{nepeivoda1982}), or a singleton formula (\cite{sasaki1986}), or a rank 0 formula (\cite{hayashi1988}) that generalize them all.} This is a well-known fact commonly used in proof extraction research to simplify extracted programs (\cite{goad1980}, \cite{sasaki1986}, \cite{berger2006}, \cite{schwichtenberg2012}).

    \item[] \textcolor{black}{\textbf{Observation 2}. The Curry-Howard correspondence is closely related to the realizability interpretation (\cite{kleene1962}, \cite{troelstra1973}, \newline \cite{schwichtenberg2012}).} 
    As far as intuitionistic propositional logic (IPC) is concerned, we can regard realizability interpretation as corresponding to the BHK interpretation, and consequently, the Curry-Howard correspondence, i.e., realizability essentially coincides with type inhabitation.
 
    \item[] \textcolor{black}{\textbf{Observation 3}. There is a type-theoretic interpretation of Kleene-Aczel's slash realizability relation in Martin-L\"{o}f's type theory that allows us to consider open proof objects computable to canonical forms (\cite{smith1993}.}
    
\end{itemize}

\medskip

\noindent \textcolor{black}{\textbf{Ad observation 3}. \cite{smith1993} translated Kleene-Aczel's slash realizability relation to a type-theoretic setting, specifically to Martin-L\"{o}f's constructive type theory, and showed that it can be used to formulate conditions for proof objects with free variables (with some specific restrictions) to be computable to  canonical forms. In other words, it allowed us to consider the computation of programs with holes, which is exactly what we need to fully justify our S rule.}

The notion of Harrop formula (under context $\Gamma$) can be carried over to a type-theoretic setting as well with a few changes necessitated by the framework: first, there should be an additional clause to cover formula variables, and, more importantly, as noted by Smith, we cannot include $\bot$ in the definition of Harrop formula/type if we want to retain the following theorem:

\medskip

\noindent \textbf{Theorem 1} (\cite{smith1993}, p. 195). If $C$ is a Harrop formula in the context $\Gamma$ then there exists a proof object $f(z)$ such  $\Gamma, z : C \mid f(z) : C$.\footnote{For simplicity, we omit the recursive clauses for the construction of the proof object $f(z)$ present in the original formulation with the exception of the clause for implication ($\to$), which we present below.}

\medskip

\noindent which we certainly want to as it is the crucial result for our construction of trivial proofs for Harrop formulas as it establishes that we can construct a proof object $f(z)$ computable to a canonical form of the formula $C$ by only using the assumption $x : C$.
From this perspective, we can say that a Harrop formula $C$ is self-constructable (analogously to self-realizable formulas from realizability interpretation, see \cite{troelstra1988}).

It is this fact (i.e., Theorem 1) that makes it possible to compute proof objects with ``Harrop-shaped holes'' in them, which brings us to the other crucial result (again due \cite{smith1993}) needed for the justification of the typed S rule: specifically, that we can have open proof objects, i.e., proof terms with free variables ranging over the Harrop formulas $C$, computable to a canonical form. This is established by the following result:

\medskip

\noindent \textbf{Corollary 1} (\cite{smith1993}, p. 196). Let $C$ be a Harrop formula and let $c(z)$ be constructed according to the corresponding clauses in Theorem 3 in \cite{smith1993} [here Theorem 1]. If $z : C \vdash b(z) : B$ then there exists a proof object in a canonical form $\mathit{can}(z)$ of the formula $B$ such that $z : C \vdash b(f(z)) = \mathit{can}(z) : B $.

\paragraph{Note.} \textcolor{black}{Observe that we do not need to consider the computational interpretation of open proof objects in general, just of those open proof objects whose free variables range over Harrop formulas and thus are computationally irrelevant, i.e., they have no computational content.\footnote{See, e.g., the literature on proof extraction, \cite{berger2006}.} Consequently, the effect of the adoption of this kind of open proof objects on notions such as canonical proofs or computation rules is limited although not insignificant. For example, the modified kind of hypothetical judgments restricted to Harrop assumptions, such as those in Corollary 1, requires the inclusion of specific forms of $\eta$-expansion (see \cite{smith1993}).}

\medskip

\textcolor{black}{Now, let us see how all the observations fit together.} 
\textcolor{black}{First, we have observed that Harrop formulas have at most one constructor (= have no computational content, Observation 1).} 
Next, we have observed that the Curry-Howard correspondence closely parallels the realizability interpretation (Observation 2). Of special interest is Kleene-Aczel's slash realizability relation since \cite{smith1993} showed that it can be carried over to a type-theoretic setting, specifically to constructive type theory, which we adopt in this paper. Most importantly for us, it can be used to specify conditions for proof objects with free variables (ranging over Harrop formulas) to be computable to a canonical form (Observation 3, using Observation 1). This is then what allows us to compute the major premise $c(z) : A \vee B$ of the S rule. More specifically, utilizing the results of \cite{smith1993}, we know that if {\small \AxiomC{$ z : C $}\noLine\UnaryInfC{$c(z) : A \vee B $}\DisplayProof} then there exists a canonical value $\mathit{can}(z)$ of the type $A \vee B$ such that {\small \textcolor{black}{\AxiomC{$ z : C $}\noLine\UnaryInfC{$c(f(z)) = \mathit{can}(z) : A \vee B .$}\DisplayProof}} And we know that $\mathit{can}(z)$ of $A \vee B$ has to be either $\textsf{inl}(a(z))$ or $\textsf{inr}(b(z))$, which is what we needed to establish for the computation rules for the selector \textsf{S} (introduced in Section \ref{sec:types}) to work as intended.

\section{\textcolor{black}{Appendix: Normalization}}
\label{sec:normal}

\cite{smith1993} proved normalization for Martin-L\"{o}f's constructive type theory (\cite{martin-lof1984}, \cite{nordstrom1990}) using a type-theoretic translation of Kleene-Aczel's's slash realizability (\cite{kleene1962}, \cite{aczel1968}). Specifically, he showed that if $a : A$ can be derived (with empty context), then $a$ can be computed to a canonical form of the type $A$.

The overall structure of the proof follows closely the standard structure of normalization proofs for typed terms using Tait's reducibility/computability method (\cite{tait1967}). In fact, as Smith himself notes, Tait's computability predicate $Comp_A(a)$ can be seen as a special case of the slash translation when the context is empty, i.e., $\mid a : A$.

The relation $\Gamma \mid t : A$ is inductively defined  by the following clauses (see \cite{smith1993}):

\begin{enumerate}

    \item $\Gamma \mid c : \Pi (A , B)$ \, if \, $\Gamma \vdash c : \Pi (A , B)$ and $\Gamma \vdash c = \lambda x.b(x) : \Pi (A , B)$ and there exists a proof object $b(x)$ such that $\Gamma, x : A \vdash b(x) : B(x)$ and for all proof objects $a$, $\Gamma \mid a : A$ implies $\Gamma \mid b(a) : B(a)$.

    \item $\Gamma \mid c : \Sigma (A, B) $ \, if \, $\Gamma \vdash c : \Sigma (A, B)$, there exist proof objects $a$ and $b$ such that $\Gamma \mid a : A$ and $\Gamma \mid b : B(a)$, and $\Gamma \vdash  c = (a,b) : \Sigma (A, B)$.

    \item $\Gamma \mid c  : A + B$ \, if \, $\Gamma \vdash c : A + B$, $\Gamma \vdash c = \textsf{inl}(a) : A + B$ for some proof object $a$ such that $\Gamma \mid a : A$ or $\Gamma \vdash c = \textsf{inr}(b) : A + B$ for some proof object $b$ such that $\Gamma \mid b : A$.

    \item $\Gamma \mid c : \textsf{Id}(A, a, b)$ \, if \, $\Gamma \vdash c : \textsf{Id}(A, a, b)$ and $\Gamma \vdash c = \textsf{refl}(a) : \textsf{Id}(A, a, b)$ and $\Gamma \mid a : A$, $\Gamma \mid b : B$, and $\Gamma \vdash a = b : A$

    \item $\Gamma \mid n : \mathbb{N}$ \, if \, $\Gamma \vdash n : \mathbb{N}$ and $\Gamma \vdash n = \overline{n} : \mathbb{N}$ for some numeral $\overline{n}$.

 \item $\Gamma \mid t : \bot$  \, does not hold for any proof object $t$.
 
\end{enumerate}

We will not reproduce here the whole normalization proof, we only show that its key ingredient Theorem 2:

\medskip

\noindent \textbf{Theorem 2.} Let $\Delta$ be a context $x_1 : D_1 , \ldots , x_m : D_m (x_1 , \ldots , x_{m-1})$ and $d_1 , \ldots , d_m$ terms such that $\Gamma \mid d_i : D_i(d_1 , \ldots , d_{i-1}), 0 < i \leq m$. Then $\Delta \vdash a(\vec{x}) : A(\vec{x})$ implies $\Gamma \mid a(\vec{d}) : A(\vec{d})$.

\medskip

\noindent where $\vec{x} = x_1 , \ldots , x_m$ and $\vec{d} = d_1 , \ldots , d_m$, holds even in Martin-L\"{o}f's constructive type theory extended with the typed S rule.

Theorem 2 together with Corollary 2:

\medskip
\noindent \textbf{Corollary 2.} If $\Gamma \mid a : A$, then there exists a canonical form $can$ of the type $A$ such that $\Gamma \vdash a = can : A$.
\medskip

\noindent which follows from the definition of slash, then give the normalization result (for details, see \cite{smith1993}).\footnote{The judgemental equality $a = can : A$ is understood as definitional equality.}

\medskip

The type-theoretic version of the S rule for full Martin-L\"{o}f's constructive type theory is as follows:
\begin{prooftree}
\def\fCenter{\ \vdash\ }
\Axiom$\Delta   , z : C(\vec{x})  \fCenter c(\vec{x}, z) : A(\vec{x},z) + B(\vec{x},z)$
\noLine
\UnaryInf$\Delta , x : C(\vec{x}) \to A(\vec{x})  \fCenter d(\vec{x}, x) : D(\vec{x}, \lambda z . \textsf{inl}(x) )$
\noLine
\UnaryInf$\Delta , y : C(\vec{x}) \to A(\vec{x})  \fCenter d(\vec{x}, y) : D(\vec{x}, \lambda z . \textsf{inr}(y) )$
\UnaryInf$\Delta \fCenter \mathsf{S}( c(\vec{x}, \textcolor{black}{z}) , d(\vec{x},\textcolor{black}{x}) , e(\vec{x},\textcolor{black}{y}) ) : D(\vec{x}, c(\vec{x}, \textcolor{black}{z}))$

\end{prooftree}

\noindent \emph{Note.} The operator $\mathsf{S}$ binds the variables $z, x, y$ in terms $c(\vec{x},z)$, $d(\vec{x},x)$, and $e(\vec{x},y)$, respectively.

\medskip

\noindent Computation rules:

\begin{itemize}
    \item split left: $ \textsf{S}(\textsf{inl}(a(z)), d, e) = d(\lambda z.a) : D(\textsf{inl}(a(z))) $

    \item split right: $ \textsf{S}(\textsf{inr}(b(z)), d, e) = e(\lambda z.b) : D(\textsf{inr}(b(z))) $

\end{itemize}

\noindent \emph{Proof of Theorem 2.} By induction on \textcolor{black}{the structure} of the derivation $\Delta \vdash a(\Vec{x}) : A(\vec{x})$ (together with lemmas 1 and 2, see below). As mentioned above, we show only the case for the new typed rule S, i.e., the selector $\mathsf{S}$. \textcolor{black}{The presentation of the proof itself follows \cite{smith1993}'s treatment of the elimination rule for disjoint union/disjunction to make the differences between selectors \textsf{S} and \textsf{D} more apparent.}

\medskip

\noindent \textcolor{black}{By induction hypothesis we obtain}:

\medskip

\begin{itemize}

\item[(1)] $\Gamma \mid c(\vec{d} , z ) : A(\vec{d}, z ) + B(\vec{d},z)$ for all $z$ such that $\Gamma  \mid z : C(\vec{d}) $

    \item[(2)] $\Gamma \mid d(\vec{d}, \lambda z . a(z)) : D(\vec{d}, \lambda z . \mathsf{inl}(a))$ for all $\lambda z . a(z)$ such that $\Gamma \mid \lambda z . a(z) : C(\vec{d}) \to A(\vec{d},z)$

    \item[(3)] $\Gamma \mid e(\vec{d}, \lambda z . b(z)) : D(\vec{d}, \lambda z . \mathsf{inr}(b))$ for all $\lambda z . b(z)$ such that $\Gamma \mid \lambda z . b(z) : C(\vec{d}) \to B(\vec{d},z)$
    
\end{itemize}

\noindent From (1) we get via the definition of slash either:

\medskip

\begin{itemize}

    \item[(4)] $\Gamma \vdash c(\vec{d}, z) = \mathsf{inl}(a(z)) : A(\vec{d},z) + B(\vec{d},z)$ for some term $a(z)$ such that $\Gamma \textcolor{black}{, z : C(\vec{d})} \mid a(z) : A(\vec{d},z)$. $\Gamma \textcolor{black}{, z : C(\vec{d})} \mid a(z) : A(\vec{d},z)$ implies $\Gamma \textcolor{black}{, z : C(\vec{d})} \vdash a(z) : A(\vec{d},z)$. Furthermore, since $C(\vec{d})$ is a Harrop formula we can obtain, via Corollary 4 (\cite{smith1993}), that $\Gamma , z : C(\vec{d}) \vdash a(c(z)) = can(z) : A(\vec{d},z)$, i.e., that $a(c(z))$ of type $A(\vec{d},z)$ can be computed to a canonical value even if it contains free variables, as long as those variables range over Harrop formulas. The term $c(z) : C$ is recursively constructed from the assumption $z : C$ as shown in \cite{smith1993}, Theorem 3.

 \item[(5)] Analogously to (4).

\end{itemize}

\noindent Let us assume that (4) holds \textcolor{black}{and continue by case analysis}. By definition of $\mathsf{split}$, we get:

\begin{itemize}
    \item[(6)] $\Gamma \vdash \mathsf{S}(\mathsf{inl}(a(z))), d(\vec{d},x), e(\vec{d},y) = d(\vec{d}, \lambda z . a(z)) : D(\vec{d}, \mathsf{inl}(a(z)))$
\end{itemize}

\noindent \textcolor{black}{Before we can put together} (2) and (4) \textcolor{black}{to obtain (7), we need to check that $\Gamma \mid \lambda z . a(z) : C(\vec{d}) \to A(\vec{d},z)$. We proceed accordingly to the definition of slash for $\Pi$ type (see above). First, (i) and (ii) are fulfilled since from (4), we have $\Gamma , z : C(\vec{d}) \vdash a(z) : A(\vec{d},z)$  and from that, via $\Pi$-intro, we get $\Gamma \vdash \lambda z . a(z) : C(\vec{d}) \to A(\vec{d},z)$.} 
As for (iii), there is a proof object $a(z)$ such that $\Gamma, z : C(\vec{d}) \vdash a(z) : A(\vec{d}, z)$, we just need to show that for all terms $c$, $\Gamma \mid c : C(\vec{d})$ implies $\Gamma \mid \textsf{ap}(a, c) : A(\vec{d}, c)$.

\medskip

\textcolor{black}{Let us begin by assuming $\Gamma \mid c : C(\vec{d})$. From this it follows via Corollary 1 (\cite{smith1993}) that $\Gamma \vdash c = can : C(\vec{d})$. Thus, we also get $\Gamma \mid can : C(\vec{d})$ (by Lemma 1).}
\textcolor{black}{Now, we want to show that $\Gamma \mid \textsf{ap}(a, c)$, i.e., $\Gamma \mid \textsf{ap}(\lambda z . a (z), c)$.
By I.H. instantiated with the help of $\Gamma \mid can : C(\vec{d})$, we get $\Gamma \mid a(can) : A$. And since we know that $\textsf{ap}(\lambda z . a (z),c) = a(c)$, we also get that $\textsf{ap}(\lambda z . a (z),c) = a(can)$.
And from that, we can finally obtain (via Lemma 1) $\Gamma \mid \textsf{ap}(\lambda z . a (z), c)$ which is we wanted to show.}

\begin{itemize}
    \item[(7)] $\Gamma \mid d(\vec{d}, \lambda z . \mathsf{inl}(a(z))) : D(\vec{d}, \lambda z . \mathsf{inl}(a(z)))$
\end{itemize}

\noindent From (6) and (7) we obtain via lemma 1:

\begin{itemize}
    \item[(8)] $\Gamma \mid \mathsf{S}(\mathsf{inl}(a(z))), d(\vec{d},x), e(\vec{d},y) : D(\vec{d}, \mathsf{inl}(a(z)))$
\end{itemize}

\noindent And from (4) and (8) and lemma 2 we get the required:

\begin{itemize}
    \item[(8)] $\Gamma \mid \mathsf{S} (c(\vec{d}, \textcolor{black}{z}),  d(\vec{d},x), e(\vec{d},y)) : D(\vec{d}, c(\vec{d}, \textcolor{black}{z}) ) $.
\end{itemize}

\bigskip

\noindent \textbf{Lemma 1} (\cite{smith1993}). Let $\Gamma \mid a : A$ and $\Gamma \vdash b : A$, then $\Gamma \vdash a = b : A$ implies $\Gamma \mid b : A$.

\medskip

\noindent \emph{Proof.} Follows from the definition of slash $\mid$. We demonstrate the case for disjunction $A \vee B$ (defined via $A + B$).

Let us assume $\Gamma \mid c : A \vee B$, $\Gamma \vdash c' : A \vee B$, and $\Gamma \vdash c = c' : A \vee B$. We want to show that $\Gamma \mid c' : A \vee B$.
By the definition of slash for disjunction, we need to show that $\Gamma \vdash c' : A \vee B$ (by assumption) and we need to show that $\Gamma \vdash c' = \mathsf{inl}(a')$ for some $a'$ such that $\Gamma \mid a' : A$ or that $\Gamma \vdash c' = \mathsf{inr}(b')$ for some $b'$ such that $\Gamma \mid b' : B$. Let us assume the first case.

Now, since we have $\Gamma \mid c : A \vee B$ (by assumption), this means we have either $\Gamma \mid a : A$ or $\Gamma \mid b :B$ (by definition of slash for disjunction), i.e., that $\Gamma \vdash c = \mathsf{inl}(a) : A \vee B$ or $\Gamma \vdash c = \mathsf{inr}(b) : A \vee B$. Let us again assume the first case. Since we have assumed that $c = c'$, this means that $\mathsf{inl}(a) = \mathsf{inl}(a')$, and thus we have found $c' = \mathsf{inl}(a)$ for $a$ such that $\Gamma \mid a : A \vee B$ and we can claim $\Gamma \mid c' : A \vee B$.

The second case proceeds analogously.

\medskip

\noindent \textbf{Lemma 2} (\cite{smith1993}). Let $\Gamma \mid a : A$ and $\Gamma \vdash B \; type$, then $\Gamma \vdash A = B$ implies $\Gamma \mid a : B$.

\medskip

\noindent \emph{Proof.} By induction on the structure of the type $A$. We demonstrate the case for disjunction $A \vee B$.

Let us assume $\Gamma \mid c : A \vee B$, $\Gamma \vdash C \vee D \; type$, and $\Gamma \vdash A \vee B = C \vee D$, and show that $\Gamma \mid c : C \vee D$.
First, $\Gamma \mid c : A \vee B$ implies $\Gamma \vdash c : A \vee B$ and from that and $\Gamma \vdash A \vee B = C \vee D$ we can conclude that $\Gamma \vdash c : C \vee D$. 
Now, we just need to show that $\Gamma \vdash c = \textsf{inl}(a) : C \vee D$ for some term $a$ such that $\Gamma \mid a : C$ or that $\Gamma \vdash c = \textsf{inr}(b) : C \vee D$ for some term $b$ such that $\Gamma \mid b : D$.
Let us consider only the first case, as the second one proceeds analogously.
From $\Gamma \mid c : A \vee B$ (by assumption) we can get (via slash definition) that either $\Gamma \vdash c = \textsf{inl}(a) : A \vee B$ for some term $a$ such that $\Gamma \mid a : A$ or that $\Gamma \vdash c = \textsf{inr}(b) : A \vee B$ for some term $b$ such that $\Gamma \mid b : B$. Let us consider the first case.
From $\Gamma \vdash c = \mathsf{inl}(a) : A \vee B$ and $\Gamma \vdash A \vee B = C \vee D$, we can conclude that $\Gamma \vdash c = \mathsf{inl}(a) : C \vee D$ for some $a$ such that $\Gamma \mid a : C$, and thus get $\Gamma \mid c : C \vee D$, which is what we wanted.

\medskip

\paragraph{Funding.}
This paper is an outcome of the project Logical Structure of Information Channels, no. 21-23610M, supported by the Czech Science Foundation and realized at the Institute of Philosophy of the Czech Academy of Sciences.

\end{document}